# First Demonstration of a Group-IV Emitter on Photonic BiCMOS Supplying a Quantum Communication Link

**Florian Honz[1], Michael Hentschel[1], Stefan Jessenig[2], Jochen Kraft[2], Philip Walther[3], and Bernhard Schrenk[1]**

[1]AIT Austrian Institute of Technology, Center for Digital Safety&Security / Security & Communication Technologies, 1210 Vienna, Austria.
[2]ams-OSRAM AG, 8141 Premstaetten, Austria.   [3]University of Vienna, Faculty of Physics, 1090 Vienna, Austria.
Author e-mail address: florian.honz@ait.ac.at

**Abstract:** We implement a silicon-on-insulator light emitter as optical supply for a QKD transmitter and transfer it to an electronic BiCMOS wafer. A secure key is established over short reach in co-existence with shortwave data transmission.      © 2024 The Author(s)

## 1. Introduction

Silicon photonics offers a rich set of functional features for a large range of applications, making it a highly attractive platform for optical communication [1], optical signal processing [2], photonic sensing [3] and quantum optics [4]. However, silicon platforms do not offer a native gain medium for light generation or amplification, unless complex hybrid or hetero-integration techniques are considered [5]. This burden can be avoided for quantum optics since the power level that needs to be provisioned to source photonic circuits, operating with single photons per symbol, is ~8 orders-of-magnitudes lower than that for classical applications. With this, the "exception" of truly monolithic silicon integration can be turned into a "rule" by accommodating light generation on silicon, in the form of – as seen from a classical perspective – inefficient, yet – as seen from a quantum perspective – powerful-enough emitters. On top of this, silicon platforms are ideally suited for electronic co-integration, for example through wafer-scale 3D integration [6]. This offers a highly attractive path to shoehorn optoelectronic quantum systems, seamlessly blended with a wide range of industrial and consumer electronics, in a very cost-efficient and non-hermetic fashion.

Towards this direction this work demonstrates, for the first time to our best knowledge, a waveguide-based group-IV light source as an optical power supply for a discrete-variable quantum key distribution (QKD) transmitter based on the polarization-encoded BB84 protocol. We transfer a 1550-nm SiGe light emitter to an electronic BiCMOS wafer and establish a secure key over a short-reach link. We further prove the robustness of the QKD link to a co-existing shortwave data channel and evaluate the limitations for joint classical/quantum communication.

## 2. Group-IV Light Emitter on Photonic BiCMOS: Fabrication and Device Characterization

In contrast to our previous work [7], which relied on a vertical silicon light emitter, this work employs a waveguide-based SiGe light source that is transferred to an electronic BiCMOS wafer (Fig. 1a). First, a 15-µm long lateral SiGe p-i-n junction is realized on a silicon-on-insulator (SOI) platform and butt-coupled to a waveguide with a 220 × 500 nm$^2$ cross-section. A 1D grating coupler has been appended to the SiGe light source for out-of-plane coupling to a standard single-mode fiber (SMF), interfacing with an external optical modulator for quantum-state preparation. The entire building block, shown in Fig. 2a, has a size of 1210 x 480 µm$^2$ and is part of a reticle that has been stepped on a 200-mm SoI wafer (*I* in Fig. 1a). This photonics wafer is then flipped and face-to-face bonded to an independently processed electronic BiCMOS wafer. The functional strata at the top layers of the SoI and BiCMOS wafers are then only separated by a thin bond oxide, permitting a bandwidth- and energy-efficient interface by means of through-oxide vias [6]. Due to the DC biasing of the silicon light emitter, this advantage can be exploited at a later stage upon monolithic integration of the optical modulator and its driving electronics. The substrate of the SoI wafer is then removed down to the buried oxide (BOX), through which the light is coupled to and from the optical waveguides of the resulting photonic BiCMOS (Φ-BiCMOS) wafer (***II***). Figure 1a further includes a cross-section of the wafer-bonding region at the Φ-BiCMOS after a focused ion beam cut.

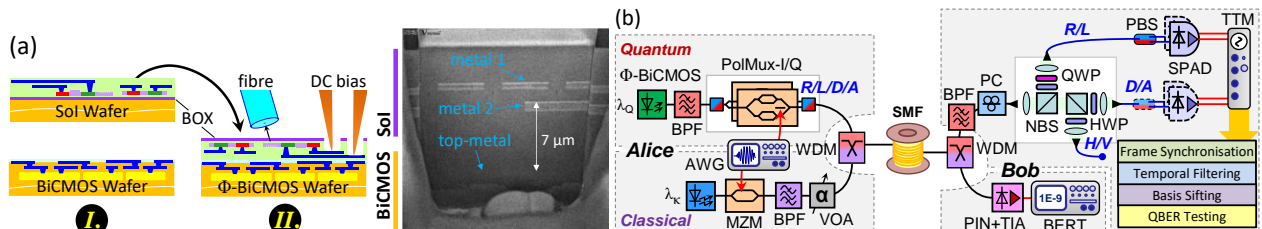

Fig. 1. (a) Fabrication of Φ-BiCMOS source and FIB cut at the bond interface. (b) Experimental setup for optically supplying a QKD transmitter.

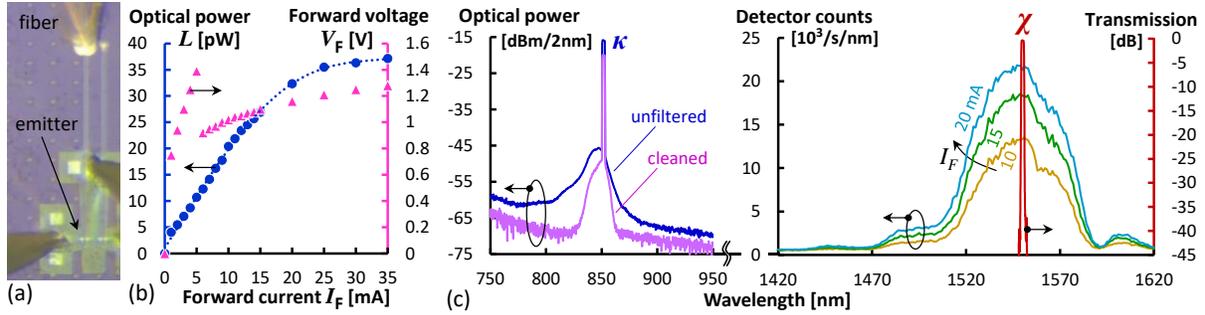

Fig. 2. (a) Φ-BiCMOS source and (b) its VLI characteristics. (c) Spectral layout for the co-existing classical (κ) and quantum (χ) channels.

The VLI characteristics of the SiGe emitter are reported in Fig. 2b. The fiber-coupled light output (●) reaches 32 µW or -74.9 dBm at a forward current of $I_F$ = 20 mA. At this bias, the spectral FWHM bandwidth of the light emission is 58 nm, centered at 1548 nm, as shown in the spectrum (Fig. 2c, χ) recorded by a single-photon spectrometer with a detection efficiency of -17 dB. The spectral characteristics of the SiGe light source align very well with the telecom-centric DWDM grid. The light output is similar to our vertically-emitting SiGe source [7], besides an earlier saturation point that limits the achievable light output. Nonetheless, we will prove that despite of the limited transmit power after optical modulation, for which the achieved values fall below the typical average photon number $\mu$ = 0.1 photons/symbol for quantum-key exchange, the waveguide-based SiGe emitter can successfully source a QKD transmitter and establish a secret key – even in co-existence with a classical channel.

## 3. Evaluation of the Silicon Light Source for Quantum Key Distribution Co-Existing with Data Transmission

Figure 1b presents the experimental setup to evaluate the Φ-BiCMOS source as an optical power supply for QKD over a short-reach link. For the *quantum channel* the SiGe emitter at $\lambda_Q$ sources a LiNbO₃ polarization-multiplexed (PolMux) I/Q modulator, which is used to encode polarization states in the right/left-circular (**R/L**) and (anti-)diagonal (**D/A**) bases [7]. This is accomplished by biasing the two unmodulated *Q* branches at their null, while retaining the two *I* branches at the maximum transmission point. When now modulating the relative phase between the two parent polarizations through driving the phase section at one of the I/Q child modulators, the four states **R**, **L**, **D** and **A** can be encoded at a constant optical power. As a pre-requisite, this traditionally DC-biased phase section has to feature an electro-optic bandwidth suitable for modulation at the quantum symbol rate $R_{sym}$. In the present case, we obtained a bandwidth of 0.92 GHz, which exceeds the requirements for $R_{sym}$ = 0.1 Gbaud. Since the output of the PolMux I/Q modulator has an average photon number of $\mu$ = 0.0148 h$\nu$/sym, there is no further attenuation required to ensure operation at a transmit level that adheres to the security requirements of discrete-variable BB84 QKD. The quantum signal is then transmitted through a 1-km link using ITU-T G.652B compatible SMF. The quantum receiver detects the signal after manual polarization control for choosing a detection basis. For the sake of simplicity, only one basis has been evaluated in one of its polarizations, thus yielding only a quarter of the possible key rate. An InGaAs single-photon avalanche photodetector (SPAD) with an efficiency of 10%, a dead time of 25 µs and dark count rate of 485 cts/s has been employed and its detection events are registered by a time tagging module (TTM) for subsequent off-line frame synchronization, temporal filtering at 50% of the symbol width and quantum bit error ratio (QBER) evaluation.

The *classical channel* employs a shortwave emitter at $\lambda_\kappa$ = 852 nm, a Mach-Zehnder modulator (MZM) and an optical PIN+TIA receiver with a responsivity of 0.56 A/W. A pseudo-random bit sequence at 1 Gb/s is transmitted at a variable launch power in order to investigate co-existence aspects. The bit error ratio (BER) is analyzed through a BER tester to investigate the safe operating area

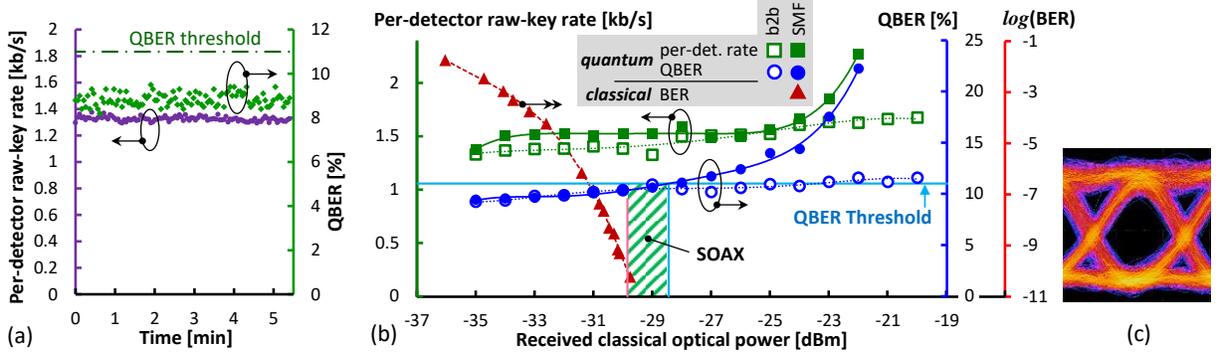

Fig. 3. (a) Back-to-back QKD performance. (b) QKD and classical performance vs classical ROP. (c) Eye for 1 Gb/s transmission at 852 nm.

for co-existence (SOAX) of the quantum and classical channels.

The spectral layout of the classical ($\kappa$) and quantum ($\chi$) channels are presented in Fig. 2c. The SiGe light emission is sliced by a flat-top 200-GHz filter centered at $\lambda_Q$ = 1550.12 nm to avoid depolarization effects over the SMF-based light path [7]. The classical channel has been allocated to the shortwave band to avoid the detrimental impact of its spontaneous emission (SE) tails and possible Raman scattering effects of an otherwise too-closely spaced adjacent classical channel. This data channel is thus located at 852 nm and is stripped off its SE tails by a 7-nm bandpass filter. The classical and quantum channels are multiplexed using standard WDM components. In order to remove any residual out-of-band noise photons at the quantum receiver, another 200-GHz filter, identical to that slicing the SiGe light emission, is employed.

## 4. Performance of Quantum Key Generation in Co-Existence with Shortwave Data Transmission

Figure 3a reports the back-to-back QKD performance without the classical data channel. A per-detector raw-key rate of 1.33 kb/s is accomplished at a QBER of 8.8%. This is below the threshold of 11% for secure-key extraction. A secure-key rate of 0.37 kb/s/detector can be yielded as the fraction of key that remains after distillation, which according to the NIST limit for fast AES key renewal (i.e., using one 256-bit AES key per data chunk of 64 GB) allows to secure a classical channel capacity of up to 745 Gb/s.

Figure 3b discusses the performance of the QKD link in co-existence with the shortwave data channel in terms of raw-key rate (□,■) and QBER (○,●) at a fixed optical budget for the quantum link while varying the transmit power of the classical channel. Results are shown for the back-to-back case (□, ○) and for transmission over a short reach of 1 km (■,●). The QBER for the back-to-back case (○) and for fiber transmission (●) are identical when the classical received optical power (ROP) level is below -30 dBm. This proves that depolarization due to the 200-GHz broad light source does not impose a limitation in combination with the fiber-based channel for short-reach links. When raising the classical power level, the QBER increases and surpasses the threshold for secure-key generation at a ROP of -23.5 dBm in case of back-to-back transmission (○). For a fiber-based link (●), this point is reached earlier at a classical ROP of -28.4 dBm. This QBER penalty and the sharp increase in detector events (■) is attributed to Raman scattering, which is known for its spectrally far-reaching noise background [8] that results in in-band noise photons for the quantum channel. The corresponding classical ROP for surpassing the QBER threshold therefore marks the upper boundary for the SOAX within which co-existence between the quantum and classical channels can be accomplished.

The classical BER transmission (▲) for the data channel at 852 nm shows a reception sensitivity of -29.9 dBm at a BER of $10^{-10}$, at which the eye diagram is clearly open (Fig. 3c). ROP levels higher than this sensitivity are required to maintain the classical data channel, therefore defining the lower boundary for the SOAX. At this ROP within the SOAX, the quantum channel would yield a secure-key rate that permits to secure a classical capacity of up to 243 Gb/s. Simultaneous quantum / classical transmission can be established, yet at a narrow dynamic range of 1.5 dB in terms of classical ROP level. The SOAX can be further extended through a more powerful silicon light emitter and a correspondingly 8.3-dB higher quantum transmit level to reach the targeted $\mu$ = 0.1 h$\nu$/sym.

## 5. Conclusion

We have demonstrated, for the first time, the feasibility of powering a QKD transmitter with a 1550-nm SiGe light source realized on Φ-BiCMOS. We accomplished a performance that is compatible with secure-key generation over a short-reach link and proved the robustness to a co-existing data channel. The use of a waveguide-based photonic platform is seen as an important leap towards full monolithic integration of a QKD transmitter, including the optical modulator. At the same time, it permits the seamless integration of electronics, which would further eliminate exposed interfaces, which are seen critical for zero-trust application environments. On top of this, the use of an all-silicon platform that omits heterogeneous III-V integration for its light source offers a path towards lowest cost when deploying quantum assets. The refinement of the light source in terms of spectral shaping and higher light output is touted to bring down the QBER while raising the secure-key rate.

Acknowledgement: This work was supported by the Austrian FFG agency and NextGeneration EU (grant no. FO999896209).